\documentclass[aps,prd,twocolumn,footinbib]{revtex4-1}
\usepackage{amsmath, mathrsfs, amssymb,amsfonts,amsthm,graphicx, epsf, dcolumn, yfonts}
\usepackage{subfigure}
\usepackage{color}
\usepackage{slashed}
\usepackage{setspace}
\usepackage{cancel}
\usepackage{epstopdf}
\usepackage[utf8]{inputenc}
\usepackage{wasysym}

\def\cblack{\color[rgb]{0.00,0.00,0.00}} 

\definecolor{cURL}{rgb}{0.00,0.00,0.62} 

\usepackage[
       colorlinks=true,
      filecolor=black,
       anchorcolor=blue,
      linkcolor=blue,
      citecolor=red, %
      urlcolor=cURL, 
       linktocpage=true,
        plainpages=false,
        breaklinks=true,
            pdfstartview=Fit 
           ]{hyperref}

\def\gh{h}
\def\ex{\mathbf{x}}

\def\q{\text{q}}
\def\tt{t}
\def\db{{\ss}}

\def\ep{{\mathbf{\epsilon}}}
\def\ve{{{\varepsilon}}}
\def\lb{\lambda_\beta}

\def\ph{{\mathbf{b}}}

\def\bd{{\Delta}}
\def\rc{{r_c}}
\def\rh{{r_h}}

\def\SL{{\mathfrak{s}}}
\def\dr{\mathrm{d}}
\def\p{\partial} 
\def\SE{{{\bf S}}}

\def\NG{_{\text{NG}}}
\def\ws{\text{ren}}

\def\fm{{\mathbf{f}}}
\def\fd{{\overset{.}{\mathbf{f}}}}
\def\fdd{{\overset{..}{\mathbf{f}}}}
\def\fddd{{\overset{...}{\mathbf{f}}}}
\def\gm{{\mathbf{g}}}
\def\gd{{\overset{.}{\mathbf{g}}}}
\def\gdd{{\overset{..}{\mathbf{g}}}}
\def\ed{{\overset{.}{\mathbf{\epsilon}}}}
\def\edd{{\overset{..}{\mathbf{\epsilon}}}}
\def\ted{{\overset{.}{\tilde{\mathbf{\epsilon}}}}}
\def\tedd{{\overset{..}{\tilde{\mathbf{\epsilon}}}}}
\def\({\left(}
\def\[{\left[}
\def\){\right)}
\def\]{\right]}

\pdfoutput=1

\begin{document}

\title{The String Worldsheet as the Holographic Dual of SYK State}
\author{Rong-Gen Cai$^{1}$}
\email{cairg@itp.ac.cn}
\author{Shan-Ming Ruan$^{2,3}$}
\email{sruan@perimeterinstitute.ca}
\author{Run-Qiu Yang$^{4}$}
\email{aqiu@kias.re.kr}
\author{Yun-Long Zhang$^{5}$ }
\email{yunlong.zhang@apctp.org}
\affiliation{\vspace{+5pt}$^{1}$CAS Key Laboratory of Theoretical Physics, Institute of Theoretical Physics,
Chinese Academy of Sciences, Beijing 100190,
and School of Physical Sciences, University of Chinese Academy of Sciences, Beijing 100049, China}
\affiliation{$^{2}$Perimeter Institute for Theoretical Physics, Waterloo, Ontario N2L 2Y5, Canada}
\affiliation{$^{3}$Department of Physics and Astronomy,
	University of Waterloo, Waterloo, ON N2L 3G1, Canada}
\affiliation{$^{4}$Quantum Universe Center, Korea Institute for Advanced Study, Seoul 130-722, Korea}
\affiliation{$^{5}$Asia Pacific Center for Theoretical Physics, Pohang 790-784, and Center for Quantum Spacetime, Sogang University, Seoul 121-742, Korea}
\vspace{+15pt}

\begin{abstract}
Recent studies of the fluctuations of an open string in AdS space show some pieces of evidence that the string with a worldsheet horizon could be a  dual description of SYK model, as they saturate universal chaos bound and share the same symmetry. An open string hangs from the AdS boundary to the horizon of black brane could be dual to a 0+1 dimensional boundary state.
To be specific, we find that the fluctuation of the string in charged BTZ black hole has an asymptotic scaling symmetry, and its Euclidean IR fixed point is  governed by the quadratic order of Schwarzian action, which is just the low energy effective theory of the SYK model.
Considering the open string  worldsheet also { has natural reparametrization symmetry},  we conjecture that
{ the action of the string worldsheet is a dual description of {SYK} state.}
\end{abstract}

\renewcommand*{\thefootnote}{\arabic{footnote}}
\setcounter{footnote}{0}


\maketitle

\allowdisplaybreaks

\section{Introduction}\label{Section1}
\vspace{-0pt}

Recent studies of  the SYK (Sachdev-Ye-Kitaev) model~\cite{PhysRevLett.70.3339} and its gravitational dual have excited the research enthusiasm in both sides of the condensed matters~\cite{PhysRevB.52.9590}
and the holographic dualities~\cite{Maldacena:2016hyu}.
The SYK model is a strongly interacting quantum system that is solvable at large N,
which leads to new insights in quantum gravity and conformal field theories.
One conjecture on the duality relies on the similarity between SYK model and the AP (Almheiri-Polchinski) model  with the near {AdS$_2$} geometry in dilation gravity
\cite{Almheiri:2014cka},  
as they share the same symmetry and  saturate the universal chaos bound.

The studies of chaos bound and  butterfly effects were initiated in
\cite{Shenker:2013yza} 
and have been generalized to various gravitational setups
\cite{Leichenauer:2014nxa}. 
In the holographic theories with Einstein gravity duals one finds that
the Lyapunov exponent $\lambda_L$ has a universal bound \cite{Maldacena:2015waa},
\begin{align}\label{Lbound}
\lambda_L\leq \lb\equiv\frac{2\pi}{\beta}.
\end{align}
The other known model which saturates the same bound is the SYK model \cite{PhysRevLett.70.3339}.
Thus, it is expected that the two dimensional dilation Einstein gravity  which contains a near AdS$_2$   sector could be a dual theory of the {SYK} model.
However, 
the exact gravitational dual formula of SYK model is still unclear \cite{Maldacena:2015waa,Jensen:2016pah}.

Two recent papers shade some new lights on the dual models of the {SYK} states~\cite{Murata:2017rbp,deBoer:2017xdk},
with an string worldsheet embedded in an higher dimensional AdS spacetime.
Originally these models are used to study the interaction between the heavy quark (or Brownian particle)
and the strongly coupled thermal plasma.
In~\cite{Murata:2017rbp,deBoer:2017xdk}, the authors firstly evaluate the Lyapunov exponent from correlation function, and find that
$\lambda_L{=}2\pi/\beta$, which saturates the universal chaos bound.
The other models  saturating this bound are the black holes and SYK states.
From these observations, we then conjecture that the action of the string worldsheet in an AdS background could be a candidate dual description of the SYK model.

In this paper,
we will show that the  $0+1$ dimensional {SYK} state can be naturally thought as the ``{SYK} quasiparticle''.
The dual description could be an open string connecting the  black brane horizon and the AdS boundary.
We will prove that
the fluctuation of the string embedded in charged BTZ black hole is dual to a one dimensional system which has an asymptotic scaling symmetry.
This leads that its {IR} fixed point  is governed by the same
{quadratic} Schwarzian action, which is the IR effective theory of {SYK} model.
Considering the open string worldsheet also has the {SL(2,R)} symmetry,  we conjecture that the {renormalized on-shel action} of the string worldsheet could be a dual description of {SYK} model.

\vspace{-0pt}
\section{The symmetries}\label{Section2}
\vspace{-0pt}

To argue two different theories  could be dual to each other,
one starting point is to confirm that they share the same symmetry.
{In the IR limit, SYK model shows emergent conformal symmetry that is explicitly and spontaneously broken into SL(2,R). The residual exact symmetry is obvious when we express its effective action as Schwarzian action } in Euclidean signature
~\cite{Maldacena:2016hyu,Stanford:2017thb},
\begin{equation}\label{actSYK1}
  \SE_{\text{Sch}}:=-\frac{1}{g_s^2} \int_0^\beta {\dr}\tau   \{\fm(\tau),\tau\}\,, \quad \frac{1}{g_s^2}\equiv\frac{\alpha_S N}{ \mathcal{J}},
\end{equation}
with the Schwarzian derivative
\begin{equation}\label{SWD}
  \{\fm(\tau),\tau\}:
  =\frac{\fddd}{\fd} - \frac{3}{2}\left(\frac{{ \fdd} }{ {\fd} }\right)^2  \,.
\end{equation}
The coupling constant $g_s$ is related to the original SYK coupling $ \mathcal{J}$, the number of Majorana fermions $N$,
and the coefficient $\alpha_S$ is determined by the even number $q$ of interacting fermions \cite{Maldacena:2016hyu}.
As its most important property, Schwarzian derivative is invariant under {SL(2,R)} transformation:
$\fm\rightarrow(a\fm+b)/(c\fm+d)$ with $ad-bc=1$. {The AP model with the near AdS$_2$ geometry as the candidate duality of SYK model also exhibits the same pattern in explicit and spontaneous symmetry breaking  \cite{Almheiri:2014cka}.}

Now let us consider an alternative dual description of SYK model,  the open string with a worldsheet horizon.
The dynamics of an open string follows from the Nambu-Goto action
of the worldsheet
\begin{equation}\label{NGAction}
S_{{\NG}} = -\frac{1}{2\pi\alpha'} \int d\sigma d\tau \sqrt{-\text{det}\,{\gh_{ab}}} \, ,
\end{equation}
where ${\gh_{ab}}=g_{\mu\nu}\partial_a X^\mu\partial_b X^\nu$ is the induced metric on the worldsheet with $a,b=\sigma,\tau$, and $X^{\mu}{(\tau,\sigma)}$ are the embedding coordinates into the target spacetime with metric $g_{\mu\nu}$.
The Nambu-Goto action is invariant under reparametrization of the worldsheet coordinates which means
\begin{equation}
\tilde{S}_{{\NG}}= -\frac{1}{2\pi\alpha'} \int d\tilde{\sigma} d\tilde{\tau} \sqrt{-\text{det}\,{\tilde{\gh}_{ab}}},
\end{equation}
under $(\sigma, \tau){\rightarrow}(\tilde{\sigma}(\sigma,\tau),\tilde{\tau}(\sigma,\tau))$. This kind of reparametrization can be apparently  considered as two-copy counterpart of conformal symmetry in SYK model and includes the {SL(2,R)} symmetry as the special case {when $\tilde{\sigma}=a\sigma+b\tau$, $\tilde{\tau}=c\sigma+d\tau$}.

{\it Small Reparametrization.} ---
Let us first consider the small reparametrization of SYK model.  If we  make the reparametrization $\fm(\tau){=}\tan\frac{\pi\tau}{\beta}{\rightarrow}\tan\frac{\pi \gm(\tau)}{\beta}$, then the Schwarzian action~\eqref{actSYK1} becomes
\begin{equation}\label{actSYK2}
  \SE_{\text{Sch}}=\frac{1}{2 g_s^2}\int _0^\beta \dr\tau\left[\left(\frac{\gdd}{\gd}\right)^2-
  \left(\frac{ 2\pi}{\beta}\right)^2  \,\big(\gd\big)^2\right]\\.
\end{equation}
Now making the small fluctuation $\gm(\tau)=\tau+\mathbf{\epsilon}(\tau)$ and expanding in $\mathbf{\epsilon}(\tau)$, we get a quadratic action,
\begin{equation}\label{actSYK3}
  \SE_{\text{Sch}}^{(2)}:=\frac{1}{2 g_s^2}\int_0^\beta\dr\tau\left[\big({\edd}\big)^2-\left(\frac{ 2\pi}{\beta}\right)^2 \,\big({\ed}\big)^2\right]\,.
\end{equation}
As we have fixed the fluctuation on the fixed parametrization $\gm(\tau)=\tau$, the quadratic action in terms of fluctuation $\mathbf{\epsilon}(\tau)$ loses the {SL(2,R)} symmetry. However, this quadratic action has an new scaling symmetry. To see this, let us first make a rescaling on time $\tau=\tilde{\tau}\mu$, then  Eq.~\eqref{actSYK3} reads,
\begin{equation}\label{actSYK3b}
\frac{1}{2 g_s^2 \mu^3}  \int_0^{\beta/\mu}\dr\tilde{\tau}\left[\big(\edd\big)^2- \left(\frac{ 2\pi}{\beta}\right)^2  \,\mu^2\big(\ed\big)^2\right]\\,
\end{equation}
which is different from its original action in Eq.~\eqref{actSYK3} and can lead a different equation of motion.
If we apply the following combinations under the rescaling,
\begin{equation}\label{newscal1}
\tilde{\tau} = {\tau}/\mu, ~~
  \tilde{\beta}(\tilde\mu)=\beta/\mu,~~
  \tilde{\ep}(\tilde{\tau})=\ep(\tau)\mu^{-3/2},
\end{equation}
with the new time $\tilde{\tau}$ and the new variable $\tilde{\mathbf{\epsilon}}(\tilde{\tau})$, the action action \eqref{actSYK3} becomes
\begin{equation}\label{actSYK3s}
 \frac{1}{2 g_s^2}\int_0^{\tilde{\beta}}\dr\tilde{\tau}
 \left[\big(\tedd\big)^2- \left(\frac{ 2\pi}{\tilde{\beta}}\right)^2  \big({\ted}\big)^2\right]  .
\end{equation}
We see that it is just as the same as the action~\eqref{actSYK3}. The transformation~\eqref{newscal1} is only the symmetry of {quadratic} Schwarzian. This is a new symmetry and is not contained in  the {SL(2,R)} symmetry. {The scaling transformation~\eqref{newscal1} shows that the conformal dimension of $\ep(\tau)$ is $3/2$. Because of this scaling symmetry, the {quadratic} Schwarzian actions of different temperatures are equivalent to each other.}

If it is ture as what we proposed, that the open string worldsheet action is a candidate dual description of {SYK} model, then its fluctuation  can also give the symmetry of Eq~\eqref{newscal1} and the dual boundary theory should be equivalent to Eq.~\eqref{actSYK3}. In the following, we will show that the fluctuation of an open string in AdS black brane is dual to a one dimensional system which has an asymptotic scaling symmetry just like  the transformation~\eqref{newscal1}.
 This symmetry leads to an {IR} theory, which is just the quadratic Schwarzian action of {SYK} shown in Eq.~\eqref{actSYK3}.

\vspace{-0pt}
\section{Action of the Worldsheet}\label{Section3}
\vspace{-0pt}
We begin with the black brane solution in $2+1$ dimensional Maxwell-Einstein gravity with a negative cosmological constant.
The generalization to higher dimensions is  straightforward.
The metric of the charged BTZ black brane is given by
\begin{align}\label{bbmetric}
ds^2 &=-r^2f(r){\dr} t^2+\frac{{\dr} r^2 }{ r^2f(r)}+r^2{\dr} x^2,
\end{align}
where
\begin{align}\label{bbmetric1}
&f(r)=1-\frac{r_h^2}{ r^2 }\left[1+{\q}^2\ln\left(\frac{r}{r_h}\right)\right] \, .
\end{align}
The horizon is located at $r=r_h$, and the Hawking temperature of the black brane is $T{=}\frac{1}{\beta}{=}\frac{(2-{\q}^2)\,{\rh}}{4\pi}$, which is identified as the temperature of the dual states on the AdS boundary $r\to\infty$. Charge parameter $q^2<2$ so that the dual temperature is large than zero.  
In the context of AdS/CFT correspondence~\cite{Aharony:1999ti}, the black brane is dual to a thermal bath on the boundary,
and an open string connecting the horizon and boundary could be interpreted as a dual particle state,
 such as the heavy quark~\cite{Herzog:2006gh} 
 or Brownian particle~\cite{deBoer:2008gu}. 
In the following we will show that the properties of the dual ``particle'' behave  as the well studied {SYK} state.

{\it Worldsheet Metric.} ---
For an open string that hangs from the AdS boundary to the horizon of the black brane,
we choose the static gauge $(\tau,\sigma)=(t,r)$ and parametrize the embedding of the string as  $X^{\mu}=\{t,r,\ex({\tt},r)\}$.
Then the position of the dual particle is given by $\ep(t)\equiv\ex(t,r_c)$, where $r_c\to \infty$ is an {UV} cut-off.
For the static particle in average $\langle \ep(t)\rangle=0$,
and the solution of the corresponding static string is $\ex(t,r)=0$.
The induced metric on the string  worldsheet  embedded in the black brane \eqref{bbmetric}
is an AdS$_2$ black hole
\begin{align} \label{inducews}
ds_{\text{ws}}^2={\gh_{ab}}{\dr}\sigma^a {\dr}\sigma^b
=-r^2f(r){\dr} t^2+\frac{{\dr}r^2}{r^2f(r)},
\end{align}
with the same $f(r)$ in Eq.\eqref{bbmetric1}.
We will consider the perturbations of the static string in the bulk,
which also induce the perturbations on top of this worldsheet metric \cite{deBoer:2017xdk}.

\vspace{-0pt}
{\it String Fluctuations}---
Now let us consider the string fluctuations in the Nambu-Goto action \eqref{NGAction}.
For simplicity, we consider the perturbation along one transverse direction,
and fix one parameterization through setting $\sigma=r$. {The fluctuation with such fixed background breaks the original reparametrization as well as the  SL(2,R)  symmetry.}
Up to the leading quadratic order of the perturbations $\ex({\tt},r)$,
it is given by 
\begin{align}\label{NGAction1}
S_{^{\NG}}&\simeq  - \frac{1}{2\pi\alpha'}\int {\dr{r}} {\dr}{\tt} \Big[1-\frac{1}{2f(r)}(\dot{\ex})^2+\frac{r^4 f(r)}2 ({\ex}^{\prime})^2\Big],
\end{align}
where $\dot{\ex}\equiv\p\ex({\tt},r)/\p{\tt}$, and ${\ex}^{\prime}\equiv\p\ex({\tt},r)/\p{r}$. This action is divergent because of the constant term in the action~\eqref{NGAction1} and the UV asymptotic behavior of $\ex({\tt},r)$. 
{Both these two divergences can be canceled by following counterterm,
%
\begin{equation}\label{count1}
  S_{\text{ct}}:=  \frac{1}{2\pi\alpha'}\int_{r=r_c}\sqrt{-\gamma}\dr {\tt}\,.
\end{equation}
$\gamma$ is the induced one dimensional metric at the cut-off boundary of the worldsheet $r=r_c$.}
We will define the renormalized on-shell action of the worldsheet as $S_{\text{\ws}} =S_{{\NG}} +S_{\text{ct}}$.
Let us make an periodic boundary condition in time $\ex(\tt,r)\sim\mathbf{x}(\tt+{\bd_0},r)$,
 we extract  the following quadric order of the Nambu-Goto action of the worldsheet
\begin{equation}\label{newact1}
S_{{\NG}}^{(2)}=-\frac1{4\pi\alpha'}\int_{\rh}^{\rc} {\dr{r}} \int_{-\frac{{\bd_0}}2}^{\frac{{\bd_0}}2}{\dr}{\tt} \left[r^4 f(r) ({\ex}')^2-\frac{1}{f(r)}(\dot{\ex})^2\right]\,.
\end{equation}
As we work in the Lorentz signature, this period has nothing  to do with the inverse temperature $1/T=\beta$. The value of $\Delta_0$ will be determined later to match the quadratic order of the Schwarzian action~\eqref{actSYK3} in Euclidean signature.
Let us make a Fourier's transformation ${\ex}({\tt},r)=\frac1{\sqrt{2\pi}}\sum_{n=-\infty}^{\infty}{\ph}_n(r)e^{i{\lambda_n{\tt}}}$
with ${\lambda_n\!:}{=}2\pi n/{\bd_0}$, then 
the renormalized quadric order action of the string worldsheet is
\begin{align}\label{WSAction2}
S_{{\ws}}^{(2)}:=S_{{\NG}}^{(2)}+S_{\text{ct}}^{(2)}= \sum_{n=-\infty}^{\infty}{\SL}_n,
\end{align}
with the $n$-th induced action,
\begin{equation}\label{valuepsi4}
\begin{split}
{\SL}_n&{=}\frac1{4\pi\alpha'}\Big\{\int_{r_h}^{r_c} {\dr{r}}\Big(\frac{\lambda_n^2}{f}{\ph}_n{\ph}_{-n}-r^4 f {\ph}'_n{\ph}'_{-n}\Big)
 \\ & -\lambda_n^2{\ph}_n{\ph}_{-n}|_{r_c}\Big\}.
\end{split}
\end{equation}
We see that ${\SL}_{-n}={\SL}_{n}$.
Since $\mathbf{x}({\tt},r)$ is real valued, we see that $\mathbf{b}_{-n}(r)=\mathbf{b}_n^*(r)$. Thus the induced action \eqref{valuepsi4} has a global $U(1)$ symmetry with the following conserved current,
\begin{equation}\label{currentJ}
  J_n(r)=ir^4f(\mathbf{b}_n\mathbf{b}'_{-n}-\mathbf{b}_{-n}\mathbf{b}'_{n}) .
\end{equation}

In order to see what fluctuation of bulk open string  $\ex(\tt,r)$ corresponds in the boundary, let us write down the equation of motion for $\ex(\tt,r)$ in terms of $\mathbf{b}_n(r)$, which reads,
\begin{equation}\label{eomf}
  \mathbf{b}_n''+\frac{(r^4f)'}{r^4f}\mathbf{b}_n'+\frac{\lambda_n^2\mathbf{b}_n}{r^4f^2}=0 .
\end{equation}
At the horizon, we impose the in-falling boundary condition,
\begin{equation}\label{ingoing}
  \mathbf{b}_n(r)=\chi_ne^{i\lambda_n r_*},~~~r_*:=\int\frac{\dr r}{r^2f}\,.
\end{equation}
$\chi_n$ is a finite constant and determined by $\mathbf{x}(\tt,r)$ at the horizon.
Putting the equation of motion \eqref{eomf} into the action~\eqref{valuepsi4} yields the formula,
\begin{equation}\label{onshell1}
\begin{split}
(4\pi\alpha'){\SL}_n&=    -   \left.\mathbf{b}_{-n}r^4f\mathbf{b}_n'\right|_{\rh}^{\rc}  - r_c  \lambda_n^2{\ph}_n{\ph}_{-n}|_{r_c}\,,
  \end{split}
\end{equation}
with the {UV} cut-off $r_c\gg r_h$.

On the other hand, from equation of motion \eqref{eomf}, the function $\mathbf{b}_n(r)$ has the following asymptotic solution at the boundary,
\begin{equation}\label{asymR1}
  \mathbf{b}_n(r)={\ve}_n\Big[\big(1+\frac{\lambda_n^2}{2r^2}\cdots\big)-\frac{\psi(\lambda_n)}{3r^3} (1+\cdots )\Big]\,.
\end{equation}
And at the horizon, the ingoing condition implies
\begin{equation}\label{ingoing2}
  \mathbf{b}_{-n}r^4f\mathbf{b}_n'|_{r_h}=ir_h^2\lambda_n|\chi_n|^2\,.
\end{equation}
In \eqref{asymR1},  the constant ${\ve}_n$ is determined by boundary value of $\mathbf{x}(\tt, r)$ in the following way,
\begin{equation}\label{}
\lim_{\rc\to \infty}\ex(\tt,r_c) =\frac{1}{\sqrt{2\pi}} \sum_{n=-\infty}^{\infty}{\ve}_n  e^{i \lambda_n \tt} .
\end{equation}
Since the dual boundary describes a heavy quark oscillating in the thermal system, so the high frequency modes will be suppressed by
$e^{-\lambda_n/T}$ for large $n$ and lower temperature $T$.
This means ${\ve}_n $ will be suppressed exponentially for large $n$.

Notice that the  renormalized on-shell action \eqref{onshell1} reads,
\begin{equation}\label{onshell2}
\begin{split}
  (4\pi\alpha' ){\SL}_n&= i r_h^2\lambda_n|\chi_n|^2- \left.\mathbf{b}_{-n}r^4f\mathbf{b}_n'\right|_{\rc}\\
 & \quad   - r_c \lambda_n^2{\ph}_n{\ph}_{-n}|_{r_c} .
  \end{split}
\end{equation}
The conserved current $J_n$ defined in Eq.~\eqref{currentJ} implies that $\text{Im}[\mathbf{b}_{-n}r^4f\mathbf{b}_n']$ is a constant so we see that
$\text{Im}[\mathbf{b}_{-n}r^4f\mathbf{b}_n']_{r=r_c}= r_h^2\lambda_n|\chi_n|^2$.
 Thus, the action \eqref{onshell2} becomes
\begin{equation}\label{onshell3}
 (4\pi\alpha'){\SL}_n=- {\ve}_{-n}B(\lambda_n){\ve}_n\,.
\end{equation}
Here $B(\lambda_n)=\text{Re}[\psi(\lambda_n)$],
and $B(\lambda_n)$ has the following expansion in terms of $\lambda_n$,
{
\begin{equation}\label{Bnlb1}
  B(\lambda_n)=c_0+c_2\lambda_n^2+c_4\lambda_n^4+\cdots .
\end{equation}
The coefficients $c_0, c_2, c_4, \cdots$ are independent of $\lambda_n$.} One can check that for any given charge ${\q}$, we always have $c_0=0$. 
Thus,
putting \eqref{onshell3} and \eqref{Bnlb1} into  \eqref{WSAction2},
we can find that  the total renormalized on-shell action of the worldsheet finally reads,
\begin{align}\label{NGAction2}
\begin{split}
S_{{\ws}}^{(2)}=-\frac{1}{4\pi\alpha'}\sum_{n=-\infty}^{\infty}{\ve}_{-n}\sum_{k=1}^{\infty}c_{2k}\lambda_n^{2k}{\ve}_n\, .
\end{split}
\end{align}
The exponential decline of ${\ve}_n$ makes the summation to be well-defined.
Using the inverse transformation of Fourier's series, which change from $\ve_n$ in phase space to
$\ve(t)$ in the real space, Eq.~\eqref{NGAction2} then becomes
\begin{align}\label{NGAction3}
\begin{split}
S_{{\ws}}^{(2)}{=}{-}\frac{1}{2 g_s^2}
\int_{0}^{{\bd_0}}\!\dr\tt\left[-M_0(\dot{{\ve}})^2+(\ddot{{\ve}})^2 +\sum_{k=3}^{\infty}\tilde{c}_{2k}\left(\frac{\dr^{k}{\ve}}{\dr\tt^{k}}\right)^2\right].
\end{split}
\end{align}
Here we have defined $1/(2\pi\alpha'c_4)={1/g_s^2}$ in order to compare with the Schwarzian action in Eq.\eqref{actSYK1}, and $M_0=-c_2/c_4$, $\tilde{c}_{2k}=c_{2k}/c_4$. We have shifted the time  by $\tt\rightarrow\tt+{\bd_0}/2$. From the ``bulk-boundary'' correspondence in holography \cite{deBoer:2008gu}, ${\ve}(t)$ is the boundary operator dual to the bulk field $\mathbf{x}(\tt, r)$.
The renormalized action~\eqref{NGAction3} governs the dynamic of dual operator. Coefficients $c_4$ and $-c_2$ are both assumed to be positive here. Later on, we will give numerical evidence that they are indeed positive when ${\q}\neq0$. 

So far,  the period ${\bd_0}$ for time $\tt$ has been assumed to be arbitrary. Now let us  make a rescaling such that $\tilde{\tt}=\tt/\mu, \tilde{{\ve}}(\tilde{\tt})={\ve}(\tt)\mu^{-\delta}$. Then the action~\eqref{NGAction3} reads,
\begin{align}\label{NGAction4}
\begin{split}
S_{{\ws}}^{(2)}=&-\frac{1}{2 g_s^2}
\mu^{2\delta-3}\int_{0}^{\Delta_0/\mu}\dr\tilde{\tt}\Big[-M_0\mu^2\left(\dot{\tilde{{\ve}}}\right)^2+\(\ddot{\tilde{{\ve}}}\)^2  \\
&+ \sum_{k=3}^{\infty}\tilde{c}_{2k}\mu^{3-2k}\left(\frac{\dr^{k}\tilde{{\ve}}}{\dr\tilde{\tt}^{k}}\right)^2\Big]\, .
\end{split}
\end{align}
We see that if
\begin{equation}\label{scaling2}
  M(\mu)=M_0\mu^2, ~~
  \bd(\mu)={\bd_0}/\mu,~~
  \delta=3/2,
\end{equation}
then the action \eqref{NGAction4} will has an asymptotic scaling invariance  when $\mu\rightarrow\infty$, which implies the following renormalization equation,
\begin{equation}\label{RGeq1}
  \frac{\dr}{\dr\mu}(M\bd^2)=0\,.
\end{equation}
Now take the initial value of ${\bd_0}$ to satisfy $M_0{\bd_0}^2=4\pi^2$,
 then at the {IR} limit ($\mu\rightarrow\infty$),
 we can drop the higher order terms, and the action~\eqref{NGAction4} reads,
\begin{align}\label{NGAction50}
S_{{\ws}}^{(2)}=&-  \frac{1}{2 g_s^2}
\int_{0}^{\bd}\dr\tilde{\tt}\left[\(\ddot{\tilde{{\ve}}}\)^2
-\left(\frac{ 2\pi }{\bd}\right)^2 \(\dot{\tilde{{\ve}}}\)^2  \right] .
\end{align}
In order to compared with the {quadratic} Schwarzian action \eqref{actSYK3} in  Euclidean signature, let's change \eqref{NGAction50} into the Euclidean signature by the replacement $\tilde{\tt}\rightarrow-i\tilde{\tau}, \Delta\rightarrow-i\db$, then we find that the Euclidean IR action reads,
\begin{align}\label{NGAction5} 
{\SE}_{\ws}^{(2)}=&\frac{1}{2 g_s^2}
\int_{0}^{\db}\dr\tilde{\tau}\left[\(\ddot{\tilde{{\ve}}}\)^2
-\left(\frac{ 2\pi }{\db}\right)^2 \(\dot{\tilde{{\ve}}}\)^2  \right] .
\end{align}
This is nothing but the {quadratic} Schwarzian action in  \eqref{actSYK3}. 
The asymptotic symmetry~\eqref{scaling2} is just the rescaling symmetry~\eqref{newscal1} if we make an identification about the boundary operator $\tilde{{\ve}}(\tilde{\tau})$ and reparametrization variable $\mathbf{\epsilon}(\tau)$.
{ The period $\db$ is not the temperature of balck hole and may be different from the period in $\eqref{actSYK3}$. However, because of the scaling symmetry, the {quadratic} Schwarzian actions are equavilent to each other for all the values of $\beta$.}

\begin{figure}[!htb]
  \centering
  \includegraphics[width=.23\textwidth]{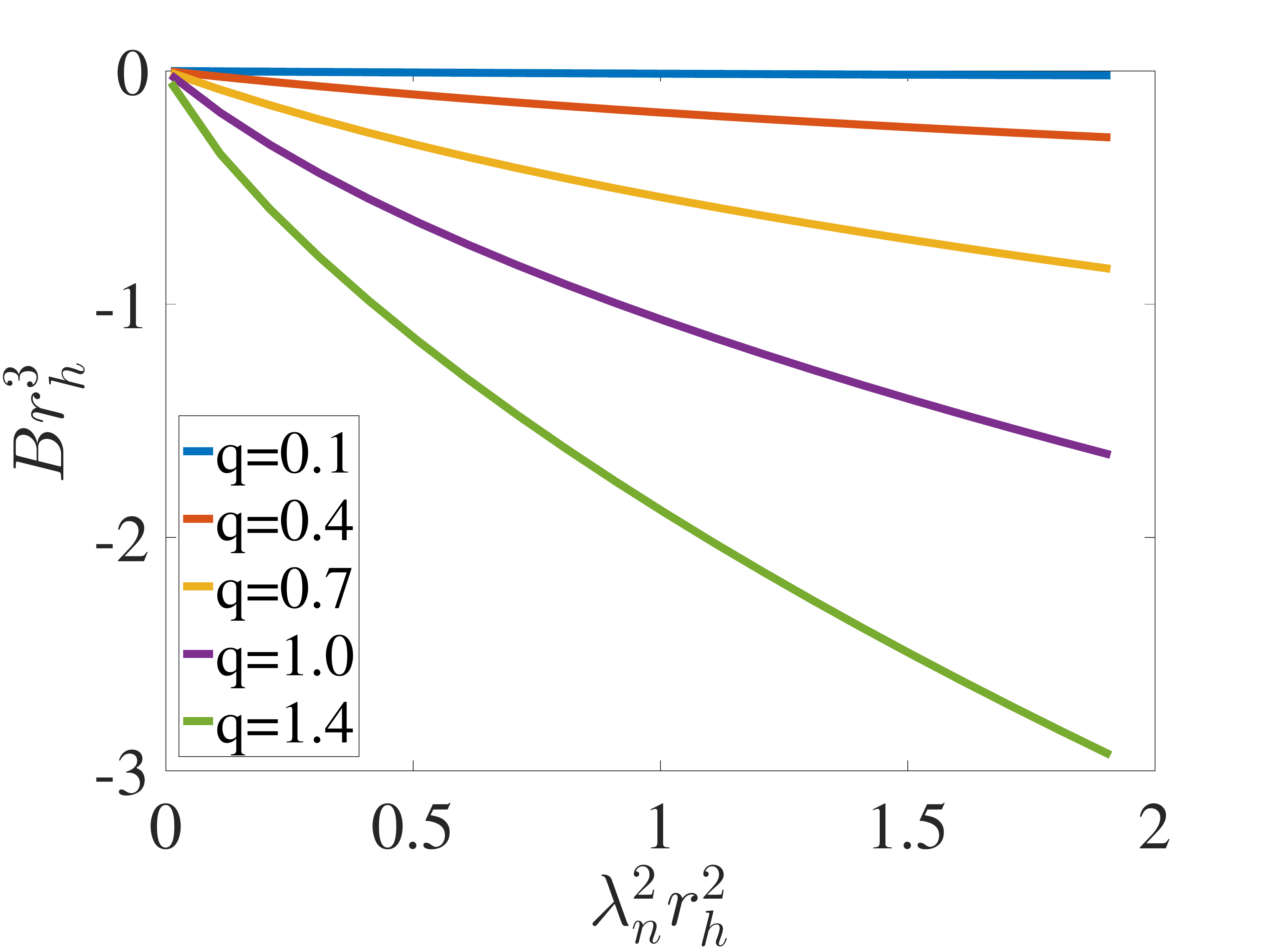}
  \includegraphics[width=.23\textwidth]{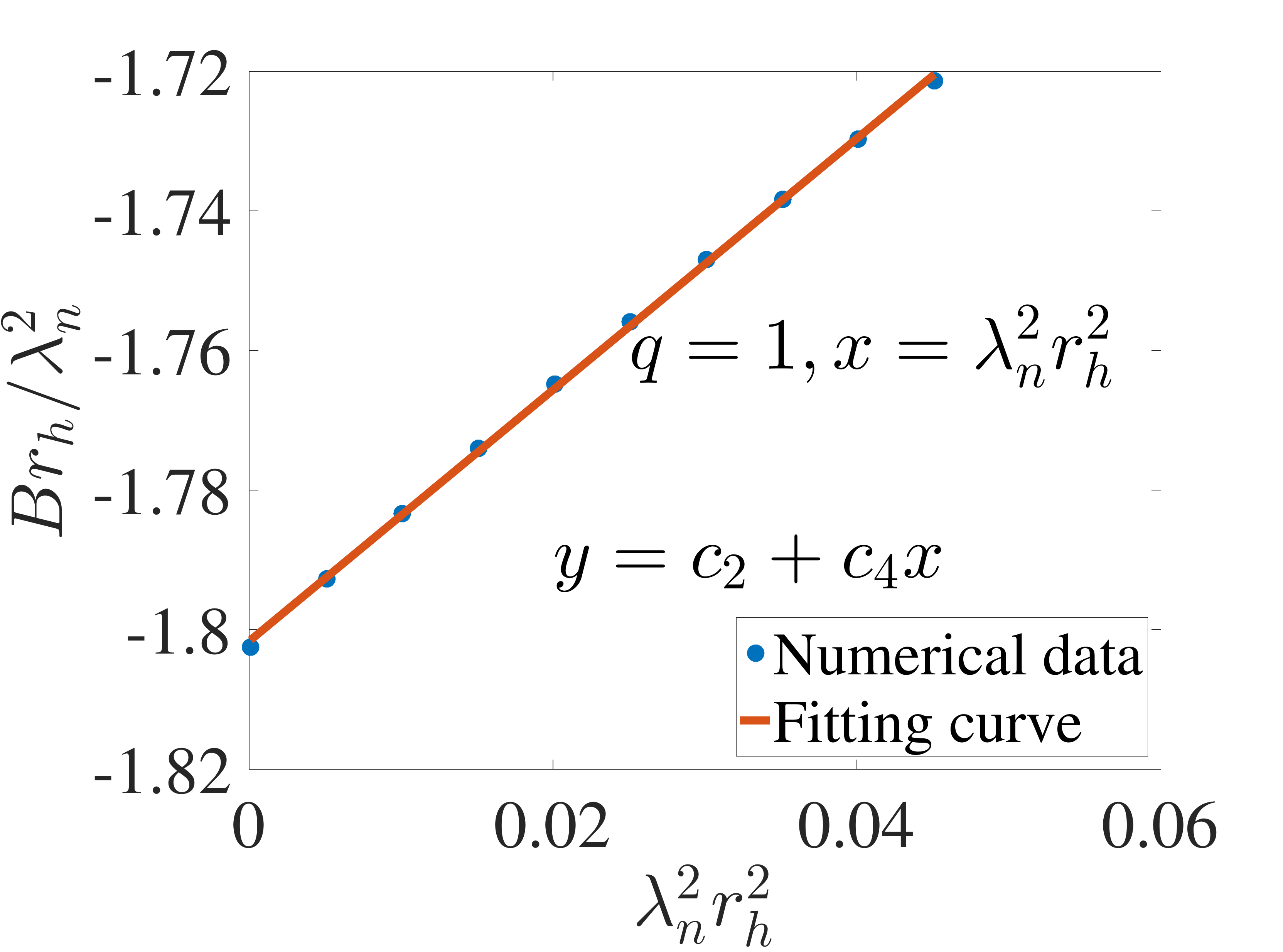}
  \caption{Left panel: Numerical values of $B(\lambda_n)$ for small $\lambda_n$ when $\text{q}=\{0.1,0.4,0.7,1.0,1.4\}$. Right panel: The fitting result for $B(\lambda_n)$ and $\lambda_n$ when ${\q}=1$, which shows that $c_2\approx-1.8$ and $c_4\approx1.8$.  } \label{Bln}
\end{figure}

Finally, we will show the numerical evidence that the coefficients $-c_2$ and $c_4$ in \eqref{NGAction2} are positive.
We first rewrite Eq.~\eqref{eomf} by the replacement $\mathbf{b}_n(r)\propto e^{i\lambda_nr_*}R_n(r)$. Then the equation of motion for $R_n(r)$ is,
\begin{equation}\label{eqforR}
  R_n''(r)+\left[\frac{(r^4f)'}{r^4f}+\frac{2i\lambda_n}{r^2f}\right]R_n'(r)+\frac{2i \lambda_n}{r^3f}R_n(r)=0\,.
\end{equation}
Because of the scaling symmetry of black brane metric \eqref{bbmetric}, we can set $r_h=1$ when solving Eq.~\eqref{eqforR} numerically. The ingoing boundary condition for $R_n(r)$ at the horizon is just he regular condition for $R_n(r)$.
Near the AdS boundary, we can just set $R_n(\infty)=1$, and the asymptotic solution for $R_n(r)$ reads,
\begin{equation}\label{asymRn}
  R_n(r)=1+\frac{i\lambda_n}r-\frac{\chi(\lambda_n)-i\lambda_n {\q}^2\ln r}{3r^3}+\mathcal{O}\Big(\frac{\ln r}{r^4}\Big).
\end{equation}
Then the value of $B(\lambda_n)$ can be expressed as,
\begin{equation}\label{relBR}
  B(\lambda_n)=\text{Re}[\chi(\lambda_n)]\,.
\end{equation}
For the case ${\q}=0$, Eq.~\eqref{eqforR} can be solved exactly and $\mathbf{b}_n(r)=e^{i\lambda_n r_*}(1+i\lambda_n/r)$,
which gives $B(\lambda_n)=0$. This means that we cannot obtain the non-trivial action in Eq.~\eqref{NGAction5} for open string worldsheet action in the neutral BTZ black hole.
Thus, we need to consider the case when ${\q}\neq0$, which can be solved numerically. The numerical results of $B(\lambda_n)$ in terms of $\lambda_n$ for different charge ${\q}$ are shown in  Fig.~\ref{Bln}.
From Eq.~\eqref{asymRn}, we see that $B(\lambda_n)$ is the even function of $\lambda_n$.
The value of $c_2$ and $c_4$ can be obtained by fitting the value of $B(\lambda_n)$ for small $\lambda_n$. For the case that ${\q}=1$, the result shows that $c_2\approx-1.8, c_4\approx1.8$. We have scand ${\q}=\{0.1,0.4,0.7,1,1.4\}$ and found that $c_2<0$ and $c_4>0$ in all cases.
%

\section{Discussions} 
\label{Section4}
Let us make a brief conclusion.
We showed nontrivial evidence that the worldsheet of an open string could be considered as the dual description of the SYK state. More precisely,  we proved that the fluctuation of an open string embedded in charged BTZ black brane is dual to a $0+1$ dimensional state which has an asymptotic scaling symmetry. This leads to the fact that its  {IR} fixed point of the Nambu-Goto action can be reduced to the {quadratic} Schwarzian action,  which is also the low energy effective action of the {SYK} model.
 Considering the open string  worldsheet also has the {SL(2,R)} symmetry,  we conjecture that the renormalized on-shel action of the string worldsheet could be a candidate dual description of the {SYK} model. In fact there exist some other pieces of evidence, which favor our conjecture.
For example, it has been shown in~\cite{Murata:2017rbp,deBoer:2017xdk}
that the  worldsheet with a horizon  saturates the universal chaos bound, as the same as the {SYK} model.
Thus, the gravity is not necessary to be included in the dual descriptions of {SYK} models.

If the dual theory of {SYK} model can be described by the string worldsheet,
it is interesting to compare it with the previous physical interpretations of the dual boundary state dual to the string in AdS, 
such as ``heavy quark''  \cite{Herzog:2006gh}, 
or ``Brown particle''.
{Brownian motion of the particle in the AdS/CFT is studied in \cite{deBoer:2008gu},
and it is quite promising to relate  the random force on the Brownian particles
with the random coupling in the {SYK} states.
Since the holographic dual of an open string in AdS can be assumed as the {SYK} quasiparticle attached at the string end point,
 the entanglement between two SYK particles is expected to be described by the holographic EPR pair~\cite{Jensen:2013ora}.
Then the well studied holographic dual model of the  EPR pair
can also be interpreted as the interaction between two ``{SYK} quasiparticles''.
Further more, multi ``{SYK} quasiparticles'' are expected to be dual to multi strings in the AdS.
In the continuous limit,  the $p+1$ dimensional ``{SYK} material'' is dual to the $(p+1)+1$ dimension brane in higher dimensional AdS,
such as the ``{SYK} chain''   or ``{SYK} layer'' \cite{Gu:2016oyy,Davison:2016ngz}.
Although the string worldsheet has the {AdS$_2$} black hole solution,
it seems also interesting to introduce the Maxwell field in higher dimensional gravity,
which is used to realize the extreme black hole  with the near horizon geometry as AdS$_2{\times}$R$_{d-1}$~\cite{Davison:2016ngz}. Thus, there are rich physics to be explored on the dynamics of open string in such a background, and it is interesting to compare them carefully with the SYK models.

\vspace{5pt}
\noindent
\emph{\bf Acknowledgements.}
We thank J.\,W.\,Chen, S.\,He, B.\,H.\,Lee, C.\,T.\,Ma, C.\, Park, Y.\,H.\,Qi, S.\,Sun for helpful conversations.
R.\,G.\,Cai was supported by the National Natural Science Foundation of China (No.11690022, No.11375247, No.11435006, No.11647601),
Strategic Priority Research Program of CAS (No.XDB23030100),
Key Research Program of Frontier Sciences of CAS.
R.\,Q.\,Yang and Y.\,L.\,Zhang benefited from the workshop ``Geometry and Holography for Quantum Criticality'' at
APCTP (Pohang, \href{https://www.apctp.org/plan.php/holography2017/1944}{\cblack{Aug.18-26, 2017}}).


%

\end{document}